\documentclass{cmspaper}
\usepackage{epsfig}
\def\pt{$p_t$}
\def\et{$E_T$}
\def\etm{$E_T^{\rm miss}$}
\def\mone{$m_{1/2}$}
\def\mzero{$m_0$}
\def\tanb{$\tan\beta$}
\def\signm{sign$(\mu)$}
\def\Azero{$A_0$}
\def\rp{$R_p$}
\def\ra{\rightarrow}
\def\Journal#1#2#3#4{{#1} {\bf #2}, #3 (#4)}
\def\CPC{\em Comput. Phys. Commun.}

\def\NPB{{\em Nucl. Phys.} B}

\newlength{\leftcolwidth}%
\newlength{\rightcolwidth}%
\newcommand{\doublecolumn}[3]%
{%
\setlength{\leftcolwidth}{#1}%
\setlength{\rightcolwidth}{\textwidth}%
\addtolength{\rightcolwidth}{-1.0\leftcolwidth}%
\addtolength{\rightcolwidth}{-2.5ex}%
\begin{minipage}{\leftcolwidth}%
{#2}
\end{minipage}%
\begin{minipage}{2.5ex}%
\mbox{\,}
\end{minipage}%
\begin{minipage}{\rightcolwidth}%
{#3}
\end{minipage}%
}

\begin{document}
\date{\today}
\title{Discovery Potential for SUGRA/SUSY at CMS}

\author{Stefano Villa \\
{\em University of California, Riverside, CA 92521, USA} }


\maketitle

\abstract{The expected SUSY discovery potential of the CMS
experiment at LHC is described, both in the MSSM and in the 
more constrained framework of mSugra, 
with emphasis on inclusive searches, 
the MSSM Higgs sector, and one example of complete
reconstruction of a SUSY decay chain.}

\vfill
\begin{center}
Presented at SUGRA20, Boston, USA,  17-21 March, 2003. 
\end{center}
\section{Introduction}
%
Searches for supersymmetry (SUSY) will be actively carried out with
the CMS detector, 
one of the two general purpose experiments which will start collecting data at the LHC
proton proton collider at CERN in the year 2007. The centre-of-mass energy  
will be 14 TeV, allowing the study of physics in the TeV range, where 
most theories predict a breakdown of the Standard Model (SM) and the appearance of
new physics. 
The LHC will probably run for some time at a luminosity of about 
$2 \times 10^{33}$~cm$^{-2}$~s$^{-1}$ (called low luminosity in the following)
before reaching the design value of $10^{34}$~cm$^{-2}$~s$^{-1}$
(called high luminosity).

The typical SUSY signatures are high transverse momentum (\pt)
jets, missing transverse energy (\etm) and possibly high-\pt \ leptons.
The experimental requirements are therefore a good jet and \etm \ resolution, 
hermetic calorimetry, efficient b tagging and tau identification and precise lepton and 
jet energy scales. 

Searches for SUSY have to deal with models with a relatively large set of free parameters, 
all consistent with the low-energy data and with constraints coming from cosmology.
The minimal supersymmetric extension of the Standard Model (MSSM) 
contains two Higgs doublets and
and has more than 100 free parameters. 
A great reduction in the number of parameters is obtained with
the unification of masses at the GUT scale and the conservation
of R parity ($R_p$). With, in addition, the requirement of a
dynamic (radiative) electroweak symmetry breaking, only five parameters
survive: $m_{1/2}$, the common 
gaugino mass at GUT scale, \mzero, the common scalar mass at GUT scale, \tanb, the ratio of the
vacuum expectation values of the two Higgs doublets, \signm, the sign of the Higgsino mixing 
parameter and \Azero, the common trilinear scalar coupling at 
GUT scale. 
This very constrained and very predictive model is called Minimal Supergravity (mSugra) and
is often chosen as a guideline to evaluate
the potentials for discovery of
new experiments, as is done in most of this note.

In the following the expected SUSY discovery potential
of the CMS detector is described. Section \ref{sec:triggers} reports results of a study of dedicated 
SUSY triggers, Section \ref{sec:inclusive} describes the expected inclusive discovery reach
in the mSugra parameter space, Section \ref{sec:Higgs} contains some results related
to the Higgs sector and Section \ref{sec:spectroscopy} describes 
the reconstruction of the full decay chain of a specific SUSY channel.

\section{SUSY Triggers in CMS}\label{sec:triggers}
%
If R parity is conserved, as is the case in mSugra, the lighest
supersymmetric particle (LSP) is stable, and the decay chains of all other sparticles
end up to the LSP. The LSP is often chosen to be the lightest neutralino or
sneutrino, weakly interacting for cosmological reasons. It therefore
escapes detection and leads to large \etm, together with many high-\pt \ jets
or leptons coming from the SUSY decay chain of the produced sparticles
(mostly squarks and gluinos at LHC). 
If \rp \ is violated, the LSP may also decay
exclusively to jets, in the case of a dominant UDD coupling considered
in the following.

The design of trigger
algorithms for an effective search for SUSY signals must therefore  
rely differently on the
multijets and \etm \ signatures.
Trigger optimizations at Level 1 (L1) 
and High Level (HLT) using GEANT simulation and fully realistic reconstruction
software have recently been performed by CMS~\cite{DAQTDR}.

The SUSY trigger has been optimised at six benchmark mSugra-inspired points, three for
the low-luminosity case and three for the high-luminosity scenario; in all cases 
\rp \ conservation and violation are considered. The parameters corresponding to
the six points are reported in Table~\ref{tab:benchmarkpts}.
\begin{table}[ht]
\begin{center}
\begin{minipage}{14.1cm}
\caption{Definition of the six mSugra benchmark points used 
for the trigger studies. The other parameters were chosen as: \Azero=0, 
\tanb=10 and $\mu > 0$. The corresponding production cross sections 
are reported in the last column.
\label{tab:benchmarkpts}}
\end{minipage}
\end{center}
\footnotesize
\begin{center}
\begin{tabular}{|c|c|c|c|}
\hline
Point      &    \mzero     &  \mone     &  $\sigma$\\
           &   (GeV/$c^2$) &(GeV/$c^2$)  &  (pb)    \\	   
\hline
4	  &     20          &   190      &  181  \\
5	  &     150         &   180      &  213  \\
6	  &     300         &   150      &  500  \\
7	  &     250         &   1050     &  0.017 \\
8	  &     900         &   930      &  0.022 \\
9	  &    1500         &   700      &  0.059  \\
\hline
\end{tabular}
\end{center}
\end{table}
The points are chosen to represent the most challenging scenarios,
from the point of view of triggering,
which CMS might face in the search for SUSY.
The first three points 
lie just above the mass reach of the
Tevatron, with relatively small sparticle masses and therefore with small
\etm \ and small jet \et. Points 7, 8 and
9 are chosen to test the ability to probe large 
sparticle masses, with correspondingly very low cross sections.

The results of the optimisation of the HLT are reported in Table~\ref{tab:trigger}, 
together with the corresponding cut.
The efficiencies achieved are satisfactory in all
points, even in the more challenging \rp-violating
configurations.
\begin{table}[t]
\begin{center}
\begin{minipage}{14.1cm}
\caption{The HLT cut values and efficiencies for six mSugra points defined in 
Table~\ref{tab:benchmarkpts} and their corresponding
\rp \ violating versions (denoted with R next to the point number). The cuts are based
on \etm  and on \et \ of the jets.
The table shows the efficiencies of the separate selections and, in parentheses, the 
cumulative efficiencies.
\label{tab:trigger}}
\end{minipage}
\end{center}
\begin{center}
\footnotesize
\renewcommand{\arraystretch}{1.1}
\begin{tabular}{|l|c|c|l|c|c|}
\hline
\multicolumn{3}{|c|}{Low luminosity} & \multicolumn{3}{|c|}{High luminosity} \\
\hline	   
Point      &  1 jet $>$ 180 GeV     &4 jets,               &Point & \etm$>$239 GeV    & 4 jets,          \\
           &  \etm$>$123 GeV     &     \et$>$113 GeV    &      &                      & \et$>$185 GeV    \\     
\hline	   
           &  efficiency (\%)       & efficiency(\%)       &      &  efficiency (\%)     & efficiency(\%)   \\ 
	   &                        & (cum. eff.)          &      &                      & (cum. eff.)      \\
\hline
4	  &   		 67	    &	 11 (69)    	   & 7    &   85	         &  18 (85) 	    \\
5	  &   		 65	    &	 14 (68)    	   & 8    &   90	         &  28 (92) 	    \\
6	  &   		 37	    &	 16 (44)    	   & 9    &   72	         &  28 (76) 	    \\
4R  	  &   		 27	    &	 28 (46)    	   & 7R   &   70 	         &  75 (90) 	    \\
5R  	  &   		 17	    &	 30 (41)    	   & 8R   &   58 	         &  78 (88) 	    \\
6R  	  &   		  9	    &	 20 (26)    	   & 9R   &   41 	         &  52 (64) 	    \\
\hline
\end{tabular}
\end{center}
\end{table}
%

\section{Inclusive SUSY Reach}\label{sec:inclusive}
%
\begin{figure}[t]
\begin{center}
\begin{minipage}{14.1cm}
\hspace*{-0.6cm}
\doublecolumn{0.55\textwidth}{
\begin{center}
\epsfxsize=16pc 
\epsfbox{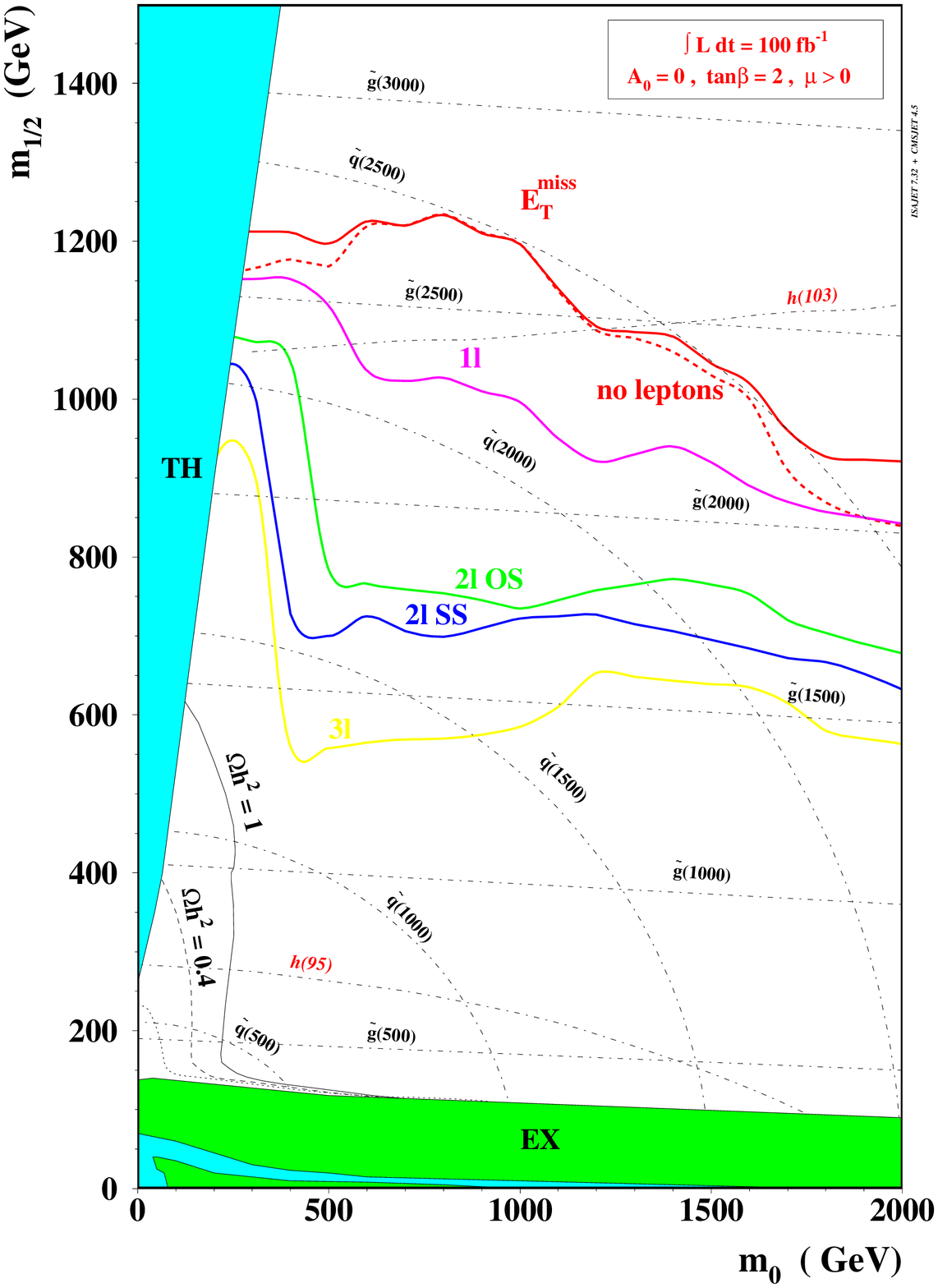} 
\end{center}
}{
\epsfxsize=16pc 
\epsfbox{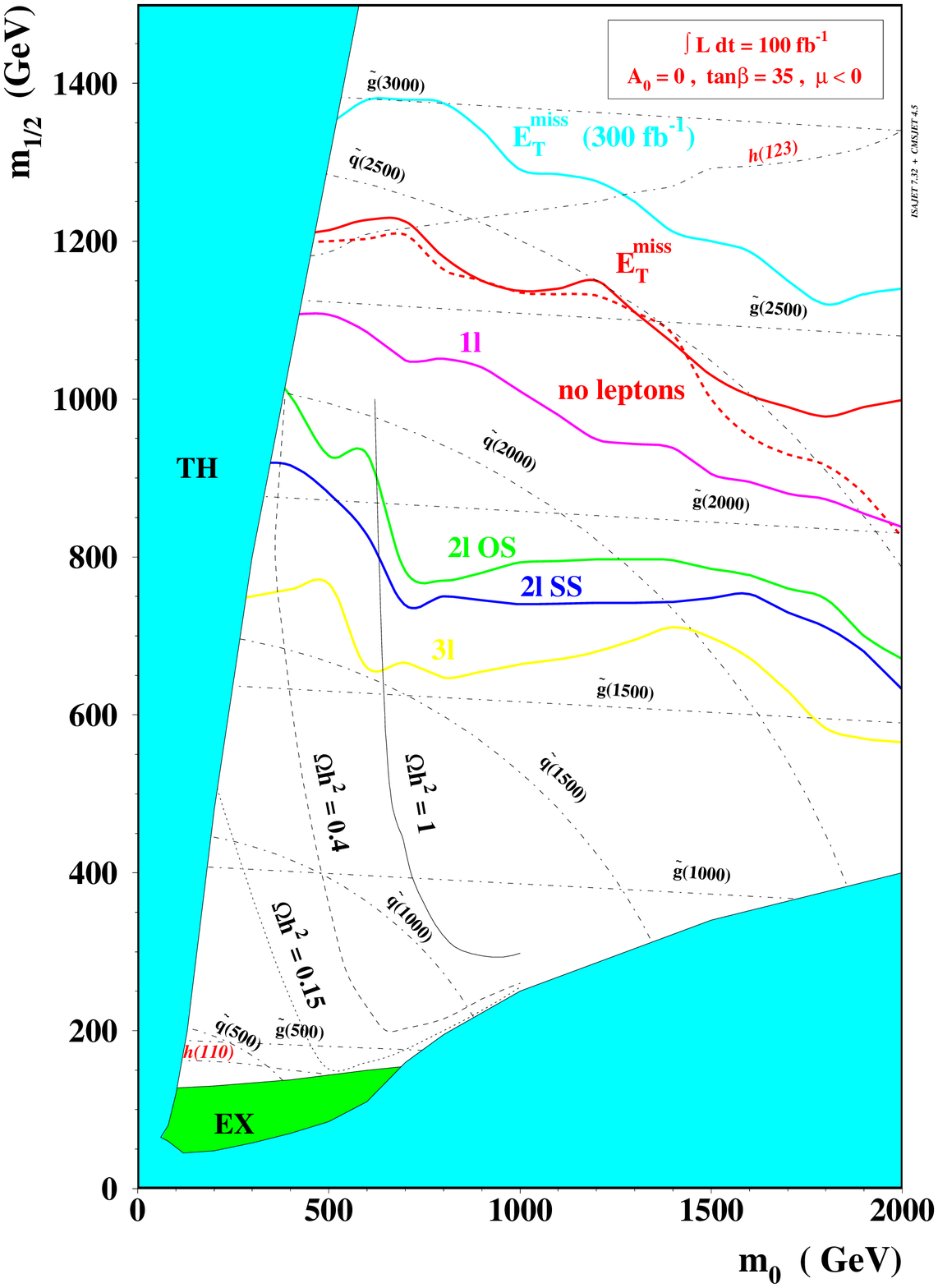} 
}
\caption{5$\sigma$ discovery reach of CMS in the mSugra \mzero-\mone \ plane 
with \Azero=0, \tanb=2, $\mu > 0$ (left) and 
\Azero=0, \tanb=35, $\mu < 0$ (right) and for 100~${\rm fb}^{-1}$ of integrated luminosity.
The full lines correspond to different final states, as defined in the text.
Dashed-dotted lines are isomass contours for squarks and gluinos. Filled areas
correspond to regions excluded either theoretically or experimentally.
\label{fig:inclusive}}
\end{minipage}
\end{center}
\end{figure}
The LHC is an ideal place to detect strongly-interacting
particles (squarks and gluon) because of the large production cross
section in proton-proton collisions.
In the \rp-conserving scenario, studies of the
inclusive discovery reach of CMS have been 
performed~\cite{abdullin}, by exploiting the already
mentioned signatures, i.e.,  missing energy, 
high-\pt \ jets from squark and gluino decays and a number of
isolated leptons, depending on the decay chains. In the case of 
high \tanb, the events also contain a large number of b quarks and $\tau$
leptons. 

Besides the common requirements of  \etm \ ($> 200$ GeV) and of at least two jets
with \et$> 40$~GeV, the following final-state topologies were investigated:
events with no leptons (0$\ell$), at least one lepton (1$\ell$), 
two opposite-charge leptons (2$\ell$OS),
two same-charge leptons (2$\ell$SS) and three leptons (3$\ell$).
The study was based on SM background samples generated with PYTHIA 5.7~\cite{PYTHIA} and
SUSY signals generated with ISAJET 7.32~\cite{ISAJET} and fast simulation of the detector
response (CMSJET 4.51~\cite{CMSJET}). 
The optimisation of the selection strategies was performed on each of the above
final-state categories and on the final state with only missing transverse energy, 
in the mSugra framework, 
for a few values of \Azero, \tanb \ and \signm.
A scan of the (\mzero, \mone) plane allows the discovery
reach of CMS to be determined. 
For example, Fig.~\ref{fig:inclusive}
shows the corresponding contours for an integrated 
luminosity of 100~${\rm fb}^{-1}$, for \Azero=0, \tanb=2 and
$\mu > 0$ (left) and \Azero=0, \tanb=35 and
$\mu < 0$ (right).

Similar results can be obtained for different integrated luminosities.
As an example in Fig.~\ref{fig:inclusive} (right) the discovery contour for the \etm \
final state for 300~${\rm fb}^{-1}$ integrated luminosity 
(expected in about three years of running at high luminosity)
is also reported. 
In summary, for most of the mSugra parameter space the ultimate discovery reach of CMS for
squarks and gluinos is between 2.6 and 3.0 TeV/$c^2$.

\section{The MSSM Higgs Sector}\label{sec:Higgs}
\begin{figure}[ht]
\begin{center}
\begin{minipage}{14.1cm}
\doublecolumn{0.5\textwidth}{
\begin{center}
\epsfxsize=15pc 
\epsfbox{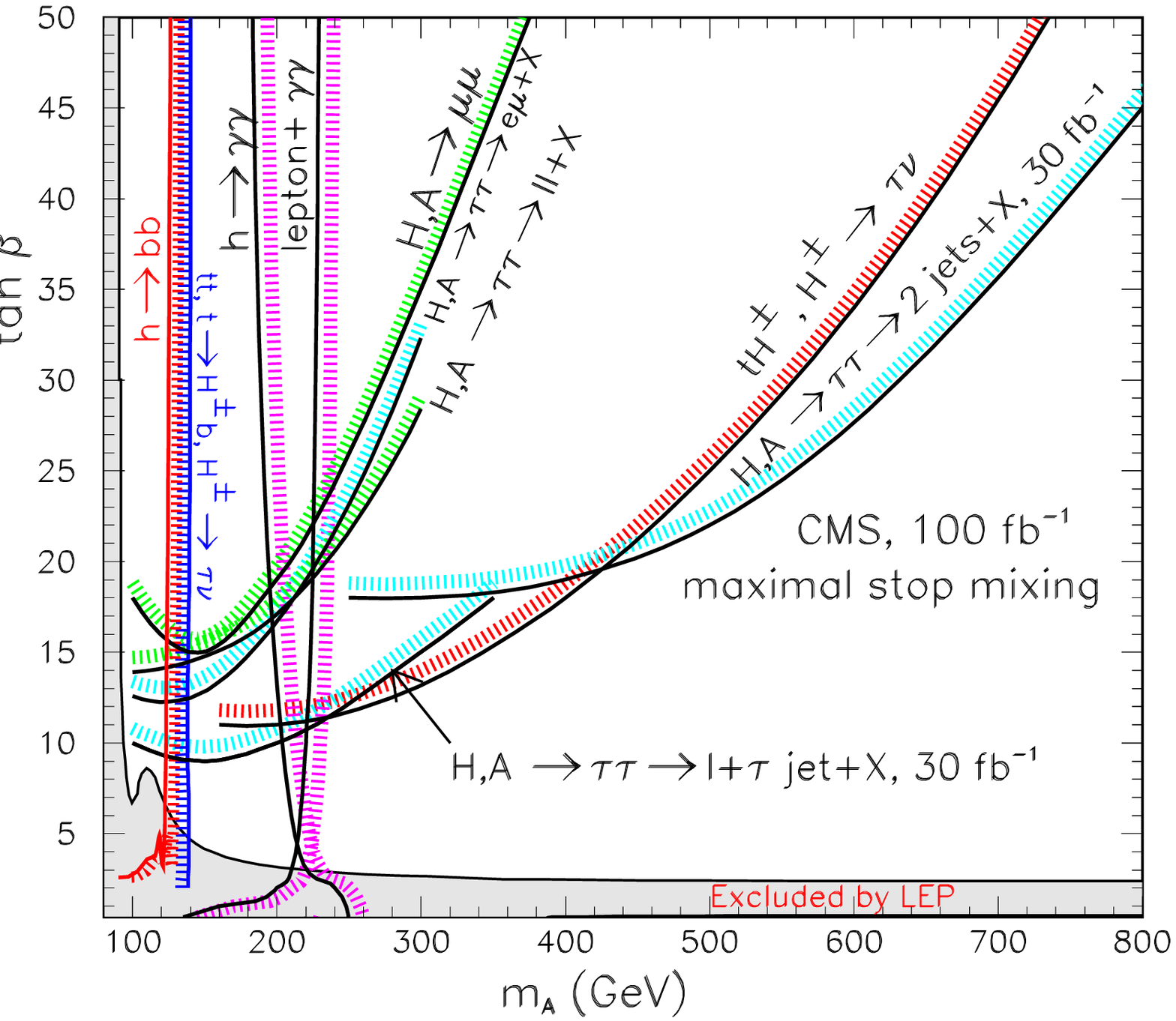} 
\end{center}
\caption{5$\sigma$ discovery contours for the MSSM  Higgs sector in CMS 
for 100~${\rm fb}^{-1}$ of integrated luminosity. 
\label{fig:Higgs}}
}{
\begin{center}
\epsfxsize=14pc 
\epsfbox{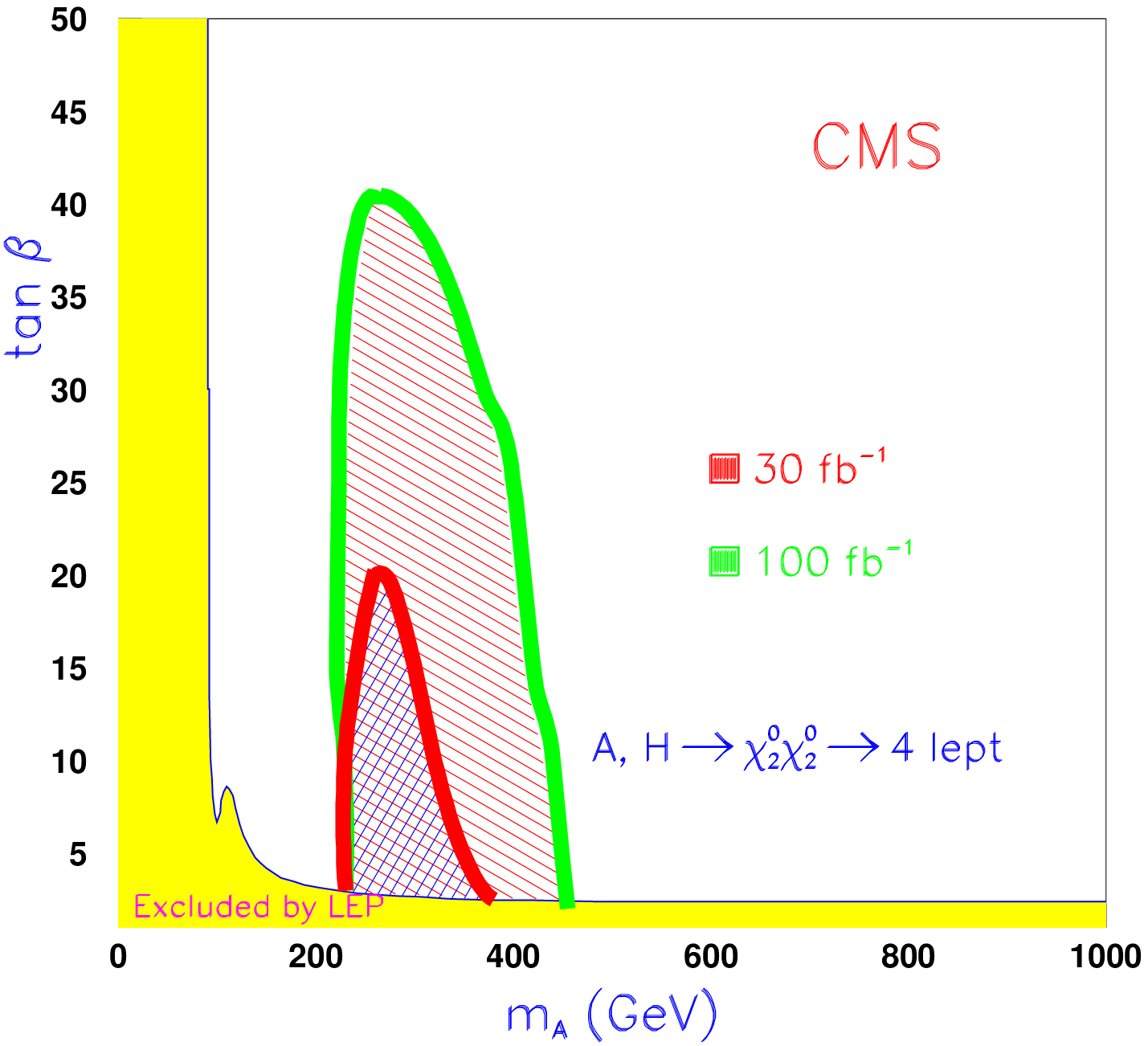}
\end{center}
\caption{5$\sigma$ discovery contours for integrated luminosity of 
30 and 100~${\rm fb}^{-1}$ for the channel A, H~$\ra \chi_2^0 \chi_2^0 \ra 4\ell + X$.
\label{fig:higgsneutr}}
}
\end{minipage}
\end{center}
\end{figure}

The Higgs sector of the MSSM consists of two SU(2) 
doublets of complex scalar fields, 
which after symmetry breaking yield 
five physical states, two CP-even bosons, h and H, one CP-odd boson A and two 
charged states H$^\pm$. 
At tree level, the Higgs sector is 
completely determined by two parameters, usually chosen to be \tanb \ 
and $m_{\rm A}$.
The discovery potential for the MSSM Higgs states is summarised in 
Fig.~\ref{fig:Higgs}, where contours of 5$\sigma$ significance are outlined
for several decay modes of the Higgs particles in the case of maximal 
stop mixing and for 100~${\rm fb}^{-1}$ of integrated luminosity. 
The lighter state, h, can
be discovered by CMS in the full parameter space.
%
%

A more challenging task is the search for the heavy states (H, A, H$^{\pm}$)
in the region
of low and moderate \tanb, 
where the couplings to the third fermion generation are not large enough
for the decays ${\rm H}^+ \to \tau\nu_\tau$ and ${\rm H, A}
\to \tau^+\tau^-$ to be exploited.
Some parts of this region can be covered by the decays of H and A to
pairs of neutralinos or charginos, when kinematically allowed~\cite{moortgat}. 
Particularly interesting is the channel A, H~$\ra \chi_2^0 \chi_2^0$
with $\chi^0_2 \ra \chi^0_1 \ell^+\ell^-$,
due to the very clear signature of four isolated leptons in the final state.
The main backgrounds are ZZ production and a few SUSY channels.
For this channel, the 5$\sigma$ discovery contours for an integrated luminosity of 
30 and 100~${\rm fb}^{-1}$ are shown in Fig.~\ref{fig:higgsneutr}, for a choice
of MSSM parameters corresponding to $M_{\chi_1^0}=60$~GeV/$c^2$ and 
$M_{\chi_2^0}=120$~GeV/$c^2$.
%
%

The production of the lightest state h in cascade decays of squarks
and gluinos, observed in the decay mode h$\ra {\rm b \bar{b}}$, 
turns out to be also a very 
promising discovery channel for this particle.
Results of a study~\cite{htobb} performed in the framework of mSugra are shown in 
Fig.~\ref{fig:htobb} as 5$\sigma$ discovery contours for integrated luminosities
of 10 and 100~${\rm fb}^{-1}$. 
The signal can be observed in a b-tagged di-jet mass distribution in 
multijet-plus-\etm \ final states, yielding in large regions of the parameter
space a signal-over-background ratio of order one. Discovery is possible
for masses of squarks and gluinos in the range from 450 GeV/$c^2$ up to about 1.5 TeV/$c^2$, 
with 100~${\rm fb}^{-1}$.

\section{SUSY Spectroscopy at CMS}\label{sec:spectroscopy}
%
Once supersymmetric particles are discovered,
it is of prime importance to study the new particles by, e.g.,
measuring their masses and branching fractions, so as to
compare with the predictions of the underlying theory.
This section describes one possible
procedure to follow in reconstructing completely a decay chain of a
sparticle in a specific example.
The framework is mSugra with \mone=250~GeV/$c^2$, \mzero=100~GeV/$c^2$, \tanb=~10, $\mu > 0$
and \Azero=0. 
The results expected with 10\,${\rm fb}^{-1}$ are shown here.

\begin{figure}[ht]
\begin{center}
\begin{minipage}{14.1cm}
\doublecolumn{0.5\textwidth}{
\hspace*{-.5cm}
\epsfxsize=16pc 
\epsfbox{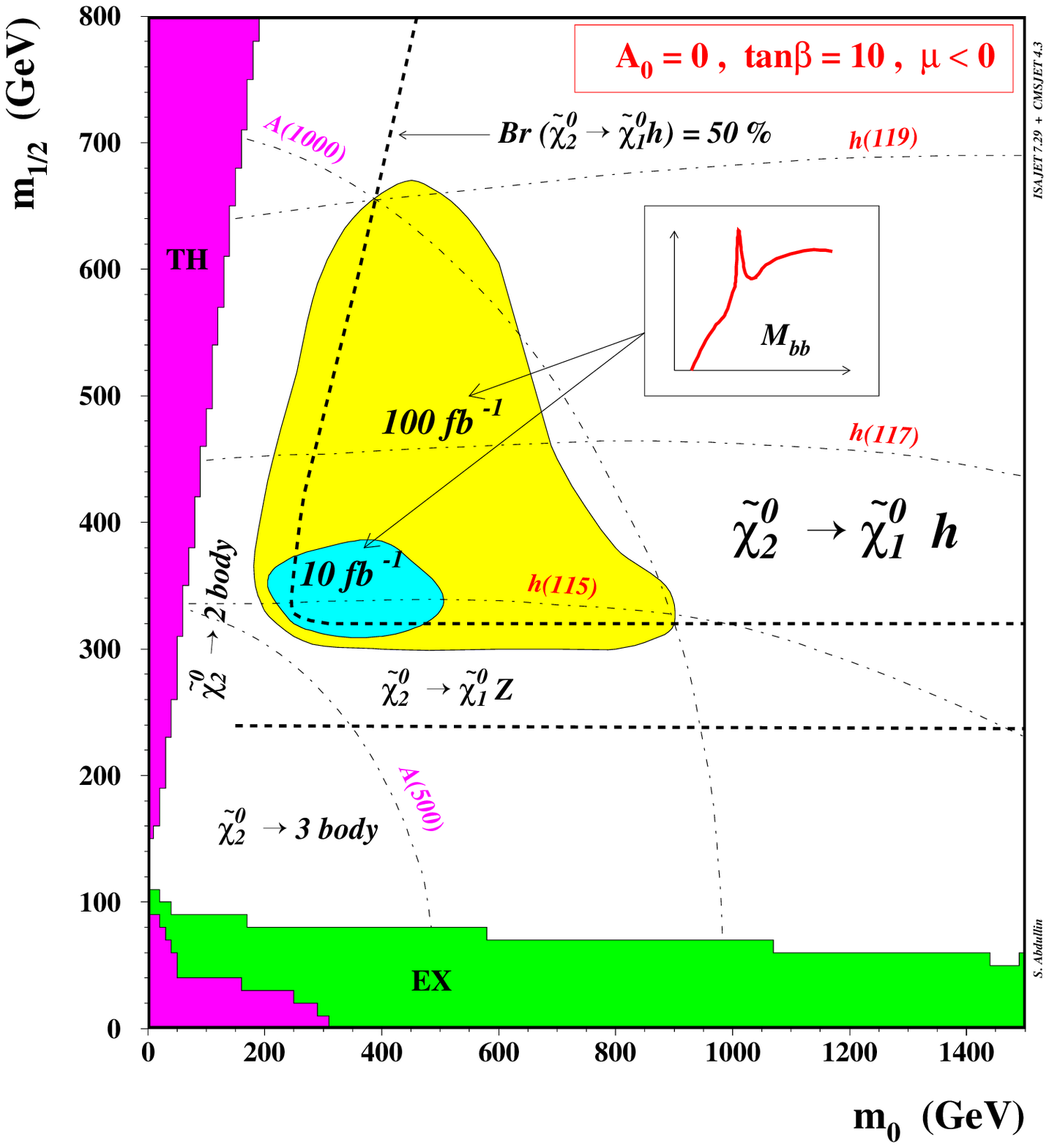} 
\caption{Discovery contours (5$\sigma$) in CMS for the lightest SUSY Higgs boson h, 
produced in SUSY cascades and decaying in the mode h$\ra {\rm b \bar{b}}$. The other 
mSugra parameters are chosen as: \Azero=0, \tanb=10 and $\mu<0$. Results are given
for integrated luminosities of 10 and 100~${\rm fb}^{-1}$.
\label{fig:htobb}}
}{
\begin{center}
\epsfxsize=14pc 
\epsfbox{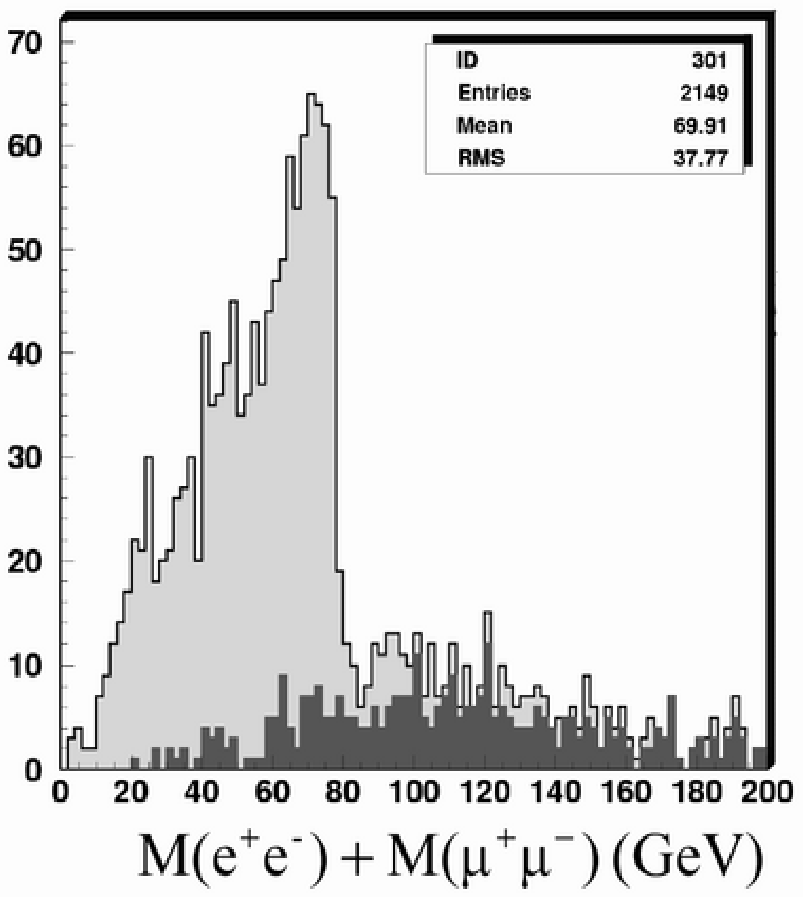}
\end{center}
\caption{The dimuon and dielectron invariant mass distribution obtained after 
requiring \etm$>150$~GeV, for the gluino-sbottom decay chain described 
in the text. The dark histogram represents the remaining SM background, the light
one is the SUSY signal.
The plot is based on an integrated luminosity of 10~${\rm fb}^{-1}$.
\label{fig:dilepton}}
}
\end{minipage}
\end{center}
\end{figure}

\begin{figure}[ht]
\begin{center}
\begin{minipage}{14.1cm}
\doublecolumn{0.5\textwidth}{
\begin{center}
\vspace*{.4cm}
\epsfxsize=15pc 
\epsfbox{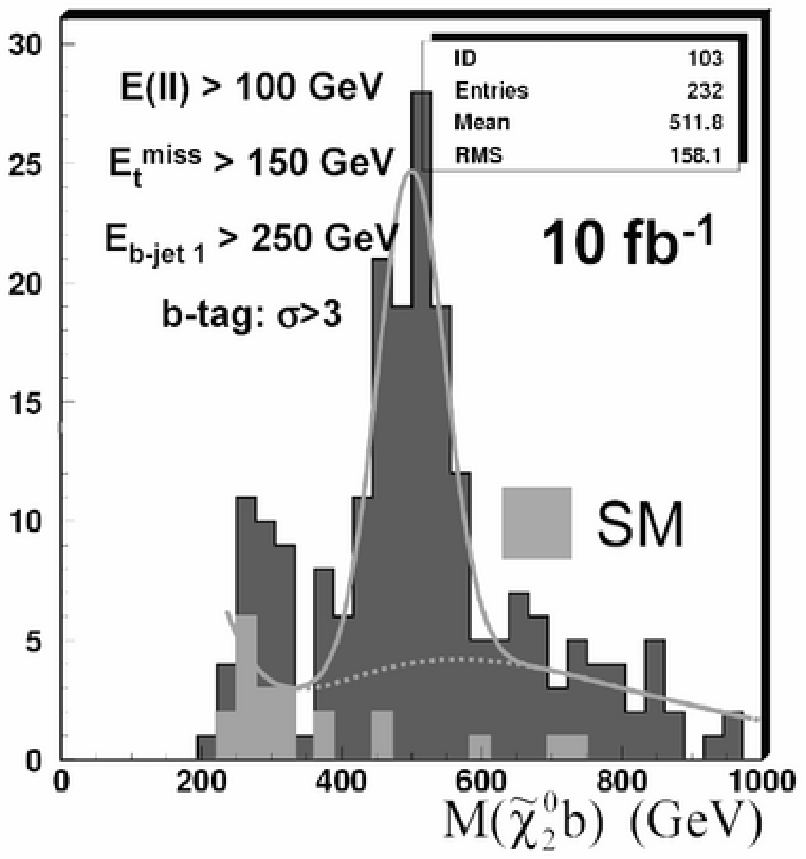} 
\end{center}
\caption{The neutralino-b invariant mass, corresponding to the sbottom reconstructed
as described in the text. The light histogram represents the SM background. Cut values
are reported as well.
\label{fig:sbottompeak}}
}{
\begin{center}
\epsfxsize=15pc 
\epsfbox{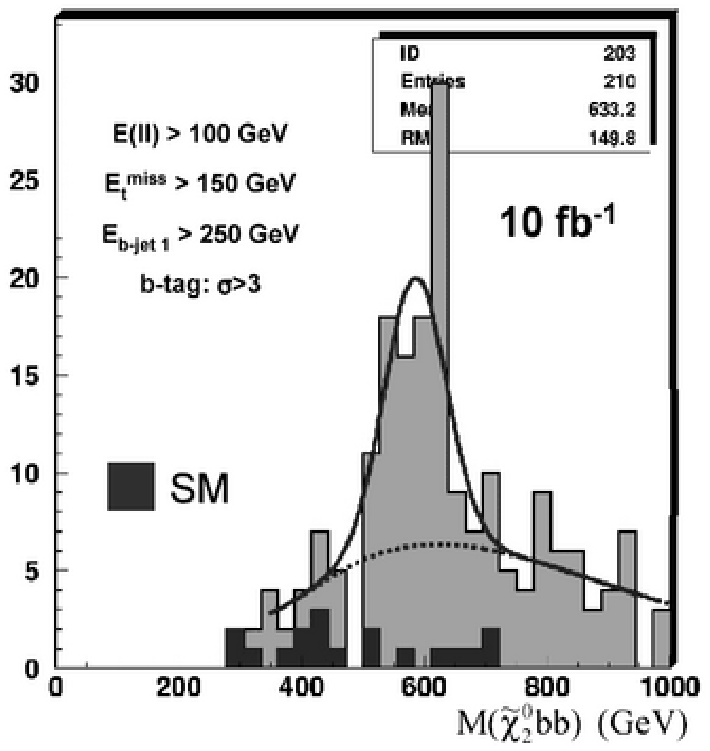}
\end{center}
\caption{The reconstructed gluino mass peak, obtained associating the sbottom and
the closest b jet. SM background is in black, SUSY signal in gray. 
\label{fig:gluinopeak}}
}
\end{minipage}
\end{center}
\end{figure}

The channel considered corresponds to the
production of a gluino, which decays through the chain 
$\tilde{\rm g} \ra \tilde{\rm b}{\rm b}$,  $\tilde{\rm b} \ra \chi_2^0 {\rm b}$, 
$\chi_2^0 \ra \tilde{\ell}^{\pm} \ell^{\mp} \ra \chi_1^0 \ell^+ \ell^-$.
The signature of this channel is the presence in the final state 
of two same-flavour opposite-sign isolated leptons (electrons and muons are considered), 
two b jets and \etm. The main SM background, ${\rm t\bar{t}}$, is greatly reduced 
by cutting on the \etm \ of the event. Figure~\ref{fig:dilepton} shows the
dilepton invariant mass distribution for events with \etm \ of at least 150~GeV.
A sharp edge due to the kinematics of the decay is clearly visible.
For events close to the kinematic endpoint, corresponding to the two
leptons being emitted back-to-back in the $\chi_2^0$  rest frame, the 
$\chi_2^0$ momentum is given by:
\begin{equation}
\vec{p}_{\chi_2^0} = \left( 1 + 
\frac{M_{\chi_1^0}}{M_{\ell^+ \ell^-}}\right)
\vec{p}_{\ell^+ \ell^-},
\end{equation}
where $M_{\chi_1^0}$ is the mass of the $\chi_1^0$, 
$M_{\ell^+ \ell^-}$ is the dilepton invariant mass and $\vec{p}_{\ell^+ \ell^-}$ is
the sum of the momenta of the two leptons.
Selecting events with a dilepton invariant mass in a window of 16~GeV/$c^2$ centred on the
dilepton edge, and associating $\vec{p}_{\chi_2^0}$ with the momentum of the
highest-\et \ b jet in the event, the kinematics of the decaying sbottom can be 
reconstructed. The sbottom invariant mass is shown in Fig.~\ref{fig:sbottompeak}.  
This distribution was obtained under the assumption that $M_{\chi_1^0}$ is known.
Alternatively, in the mSugra framework, $M_{\chi_1^0}$ can be
approximated to $M_{\chi_2^0}$/2 with little error.
Figure~\ref{fig:sbottompeak} shows that already with 10~${\rm fb}^{-1}$ the sbottom
peak is very clearly visible over a small residual background. A fit of the
peak yields a value of the sbottom mass in good agreement with the generated
value, with a resolution better than 10\%.

The final step of the analysis consists in associating the sbottom to the
closest b jet to reconstruct the gluino. The result is shown in 
Fig.~\ref{fig:gluinopeak}. The mass of the gluino is correctly extracted from the 
fit to the peak, again with a resolution of about 10\%.
The large systematic uncertainties arising from
the approximation about the $\chi_1^0$ mass remain to be studied.
%

\section{Conclusions}
%
Supersymmetry is
a good candidate theory to describe
physics at energies in the TeV range. 
The LHC therefore provides a
unique opportunity to test its predictions.
Detailed studies have been performed in CMS to evaluate the
SUSY discovery potentials in the mSugra framework.
The inclusive CMS SUSY reach, with an integrated luminosity of 300~${\rm fb}^{-1}$,
will be of about 2.6-3.0 TeV/$c^2$ for squarks and gluinos,
quite independently of the choice of parameters.
The discovery potential in the MSSM Higgs sector has also been studied.
An analysis of a gluino-sbottom decay chain in a specific
point of the MSSM parameter space shows that the 
masses of squarks and gluinos can be measured with integrated luminosities
of about 10~${\rm fb}^{-1}$ provided that the LSP mass is known.

\section*{Acknowledgments}
%
I wish to thank M. Chiorboli, F. Moortgat and S. Abdullin for providing most
of the material and D. Denegri for his help and suggestions.


\end{document}